\documentclass[twocolumn, switch]{article} 

\usepackage{preprint}
\usepackage{authblk}
\usepackage{amsmath, amsthm, amssymb, amsfonts}
\usepackage{algorithmic}
\usepackage{hyperref}
\usepackage[ruled]{algorithm2e}

\usepackage[numbers,square]{natbib}

\usepackage[utf8]{inputenc}	
\usepackage[T1]{fontenc}	
\usepackage{booktabs} 		
\usepackage{nicefrac}		
\usepackage{microtype}		
\usepackage{lineno}		
\usepackage{float}			

\usepackage{lipsum}		

\usepackage{newfloat}
\DeclareFloatingEnvironment[name={Supplementary Figure}]{suppfigure}
\usepackage{sidecap}
\sidecaptionvpos{figure}{c}

\usepackage{titlesec}
\titlespacing\section{0pt}{12pt plus 3pt minus 3pt}{1pt plus 1pt minus 1pt}
\titlespacing\subsection{0pt}{10pt plus 3pt minus 3pt}{1pt plus 1pt minus 1pt}
\titlespacing\subsubsection{0pt}{8pt plus 3pt minus 3pt}{1pt plus 1pt minus 1pt}

\definecolor{lime}{HTML}{A6CE39}
\DeclareRobustCommand{\orcidicon}{
	\begin{tikzpicture}
	\draw[lime, fill=lime] (0,0)
	circle [radius=0.16]
	node[white] {{\fontfamily{qag}\selectfont \tiny ID}};
	\draw[white, fill=white] (-0.0625,0.095)
	circle [radius=0.007];
	\end{tikzpicture}
	\hspace{-2mm}
}
\foreach \x in {A, ..., Z}{\expandafter\xdef\csname orcid\x\endcsname{\noexpand\href{https://orcid.org/\csname orcidauthor\x\endcsname}
			{\noexpand\orcidicon}}
}

\usepackage{xcolor}

\usepackage{csquotes}
\MakeOuterQuote{"}
\usepackage{xspace}
\newcommand{\mytitle}[1]{\smallskip\noindent\textbf{#1.}\xspace}
\usepackage{graphicx}
\usepackage[toc,page]{appendix}
\usepackage{subcaption}

\title{APEX: Asynchronous Parallel CPU-GPU Execution for Online LLM Inference on Constrained GPUs}

\author[1]{Jiakun Fan}
\author[2]{Yanglin Zhang}
\author[1]{Xiangchen Li}
\author[1]{Dimitrios S. Nikolopoulos}

\affil[1]{Department of Computer Science, Virginia Tech}
\affil[2]{No Affiliation}

\begin{document}

\twocolumn[ 
\begin{@twocolumnfalse} 
\maketitle
\begin{abstract}
Deploying large language models (LLMs) for online inference is often constrained by limited GPU memory, primarily due to the growing KV cache during autoregressive decoding. Hybrid GPU--CPU execution can relieve this pressure by offloading attention to the CPU, yet a key bottleneck remains: existing schedulers fail to effectively overlap CPU-offloaded work with GPU execution during the latency-critical, bandwidth-bound decode phase. This shortcoming especially penalizes real-time, decode-heavy applications (e.g., chat, Chain-of-Thought reasoning) under the memory pressure common in edge or low-cost deployments.

We present \textbf{APEX}, a profiling-informed scheduling strategy that maximizes CPU--GPU parallelism for hybrid LLM inference. Unlike systems driven by static rules or ad-hoc heuristics, APEX predicts the execution times of CPU and GPU subtasks and dynamically dispatches them to maximize overlap while keeping runtime overheads low. Critically, APEX is the first system to enable fine-grained runtime CPU offloading \emph{without batch splitting}, avoiding a common compromise that hinders performance in prior hybrid designs. APEX targets workload patterns typical of multi-turn conversation and reasoning, where prefill is brief but decode is long and performance-critical.

We evaluate APEX on diverse workloads and GPUs (NVIDIA T4, A10) using \textbf{Llama-2-7B} and \textbf{Llama-3.1-8B}. Relative to GPU-only schedulers such as vLLM, APEX improves throughput by \textbf{84\%--96\%} on T4 and \textbf{11\%--89\%} on A10, while preserving latency. Against the strongest existing hybrid schedulers, it delivers up to \textbf{72\%} (T4) and \textbf{37\%} (A10) higher throughput in long-output settings. APEX advances hybrid LLM inference on memory-constrained hardware and offers a blueprint for scheduling on heterogeneous AI systems, filling a critical gap for efficient real-time LLM applications.
\end{abstract}
\vspace{0.35cm}
\end{@twocolumnfalse}]



\section{Introduction}
The increasing scale and capability of Large Language Models (LLMs)\cite{NEURIPS2020_1457c0d6, touvron2023llamaopenefficientfoundation, zhang2022optopenpretrainedtransformer, touvron2023llama2openfoundation, grattafiori2024llama3herdmodels} have driven major advances across diverse applications. Deploying these models, however—particularly state-of-the-art variants—imposes substantial demands on underlying hardware. Moreover, eliciting sophisticated reasoning through techniques such as Chain-of-Thought\cite{wei2023chainofthoughtpromptingelicitsreasoning, kojima2023largelanguagemodelszeroshot} often requires generating extended output sequences. While GPUs provide the high computational throughput essential for LLMs, producing such sequences through autoregressive decoding involves many iterative steps. Although each decode step is typically less compute-intensive than the highly parallel prefill stage, the cumulative cost across long sequences becomes significant. More critically for system constraints, this iterative process yields a large and ever-growing KV cache for the attention mechanism. Because the cache grows linearly with sequence length, its storage emerges as a primary bottleneck~\cite{he2024fastdecodehighthroughputgpuefficientllm, kwon2023efficient, pope2022efficientlyscalingtransformerinference}.

The limited and costly onboard memory (VRAM) of GPUs presents a critical bottleneck for accommodating the KV cache. LLM engines maintain this cache in GPU memory to reuse past computations in the self-attention layer, but its substantial footprint directly constrains batch size. Although parameter sharing across multiple batched requests can raise arithmetic intensity for linear layers (e.g., Q, K, V, O projections and feed-forward networks), the KV cache frequently becomes the dominant factor limiting GPU utilization and throughput—particularly on memory-constrained devices. As a result, equipping servers with high-capacity GPUs to host large models and their expanding KV caches is prohibitively expensive, while maximizing utilization through batching only amplifies GPU memory pressure~\cite{kwon2023efficient}.

The tension between growing KV cache demands and limited GPU memory creates an opportunity to rethink the division of labor between CPUs and GPUs during LLM inference. Modern servers can be provisioned with terabytes of DRAM, motivating hybrid CPU--GPU execution. Prior work has explored offloading model weights, the KV cache, or even computation to the CPU~\cite{sheng2023flexgenhighthroughputgenerativeinference, kwon2023efficient, zhao2024hetegenheterogeneousparallelinference, song2023powerinfer, xue2024powerinfer2fastlargelanguage, 10.1145/3688351.3689164}. Early memory-offloading approaches targeted less frequently accessed data, such as model parameters or parts of the KV cache, transferring them back when needed~\cite{sheng2023flexgenhighthroughputgenerativeinference, kwon2023efficient}. However, because the KV cache is touched at every decoding step, frequent transfers over the relatively slow CPU--GPU interconnect (e.g., PCIe) become a major bottleneck, rendering techniques that rely on layer-by-layer swapping unsuitable for latency-sensitive online inference~\cite{sheng2023flexgenhighthroughputgenerativeinference, he2024fastdecodehighthroughputgpuefficientllm}.

Recognizing this challenge, recent work~\cite{he2024fastdecodehighthroughputgpuefficientllm, jiang2024neosavinggpumemory} proposes a more fundamental shift: decomposing the transformer’s execution phases and running the inherently memory-bound attention operations (which involve intensive KV cache access) directly on CPUs, leveraging their attached high-capacity DRAM. While compute-heavy components such as MLP layers remain on the GPU, this partitioning positions the CPU as an active compute participant for memory-intensive tasks, executing them ``near-memory'' and avoiding repetitive, costly GPU transfers.

This approach promises to free substantial GPU memory, enabling larger batch sizes, improving GPU utilization, and boosting overall throughput—particularly under tight memory constraints. Yet translating this concept into practical, efficient systems poses major hurdles. Achieving strong performance with CPU-offloaded attention requires careful coordination of heterogeneous resources and precise scheduling to maximize parallelism, areas where existing methods have faced practical difficulties. These challenges underscore the need for more advanced scheduling and resource management to fully realize the benefits of CPU-offloaded attention without incurring prohibitive costs or new bottlenecks.

To address these challenges, we propose \textbf{APEX}, an LLM inference system designed to asynchronously overlap CPU and GPU computation. Our primary contributions are:
\begin{enumerate}
    \item We introduce an \emph{Asynchronous Overlap Execution} mechanism that schedules CPU-based attention alongside GPU computation, reducing pipeline stalls while sustaining GPU utilization.
    \item We develop a decision mechanism guided by an empirically derived inequality (Section~\ref{sec:analytical_model_for_scheduler}) to selectively apply the optimal CPU--GPU overlap strategy and maximize throughput.
    \item We integrate these ideas into a performance-model-driven scheduler that adapts to workload and hardware characteristics based on offline profiling, delivering high throughput under memory constraints.
\end{enumerate}
By treating the CPU as an active compute tier in the inference pipeline, APEX establishes a new systems-level abstraction for heterogeneous, memory-aware scheduling that directly addresses runtime orchestration and resource efficiency. To our knowledge, APEX is the first system to enable fine-grained runtime CPU offloading without batch splitting, overcoming a key limitation of prior hybrid designs.

Experiments on T4 and A10 GPUs show that APEX improves throughput by up to 96\% over vLLM and 72\% over NEO, while preserving latency.
\section{Background and Motivation}
\label{sec:background_motivation}
\subsection{LLM Inference: Memory Pressure from KV Caches}
\label{subsec:memory_pressure}
During autoregressive generation in transformer-based LLMs, Key and Value vectors are computed and cached at each layer to avoid redundant computation. As the sequence length $N$ grows, the KV cache size increases linearly (proportional to $N \times hidden\_dims \times num\_layers$), consuming substantial GPU memory~\cite{pope2022efficientlyscalingtransformerinference, kwon2023efficient}. For example, LLaMa-65B running on four A100 GPUs can generate KV cache at 13.9~GiB/s, rapidly exhausting GPU memory~\cite{cached-attention}. This footprint frequently becomes the primary constraint on batch size, limiting overall throughput and GPU utilization~\cite{kwon2023efficient, loongserver}.

LLM inference proceeds in two phases: \textit{prefill} (processing the initial prompt) and \textit{decode} (generating subsequent tokens). The decode phase, which often dominates runtime for long sequences, is especially sensitive to KV cache management. Optimizations such as FlashAttention~\cite{dao2022flashattentionfastmemoryefficientexact} improve computational efficiency by reducing memory traffic, but they do not address the fundamental GPU memory capacity bottleneck imposed by the KV cache itself.
\begin{figure*}[]
\centering
    \includegraphics[width=\linewidth] 
    {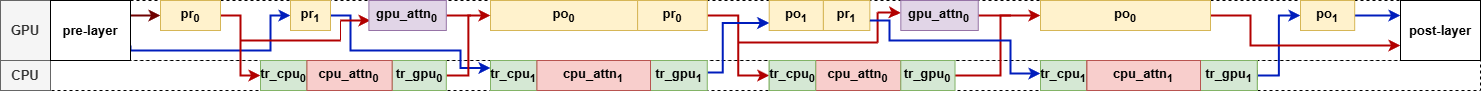}
    \caption{Asymmetric pipelining split the requests into two sub-batches. The first sub-batch (red arrows) contains prefilling and GPU/CPU decoding requests. The second sub-batch (blue arrows) contains only CPU decoding requests. “pr” means pre-projection, while "po" means post-projection + FFN operations; "attn" means attention operations; "tr" means transfer of intermediate value to GPU/CPU.}
    \label{fig:asymmetric_pipeline} 
\end{figure*}

\subsection{Hybrid CPU--GPU Inference: Evolution and Limitations}
\label{subsec:hybrid_inference_background}

The disparity between limited GPU memory and abundant host DRAM motivates hybrid CPU--GPU inference systems. Early approaches moved entire KV caches to CPU memory~\cite{sheng2023flexgenhighthroughputgenerativeinference}, but frequent swapping over slow CPU--GPU interconnects incurred prohibitive latency. A more promising direction offloads decode-phase self-attention and KV cache management to the CPU~\cite{he2024fastdecodehighthroughputgpuefficientllm, jiang2024neosavinggpumemory}, leveraging the memory-bandwidth-bound nature of decoding attention. With modern server CPUs offering $\sim$200~GB/s compared to $\sim$600~GB/s for NVIDIA A10G GPUs~\cite{he2024fastdecodehighthroughputgpuefficientllm}, this narrower bandwidth gap makes CPU execution viable while freeing GPU memory for larger batches.

FastDecode~\cite{he2024fastdecodehighthroughputgpuefficientllm} was among the first to demonstrate CPU execution of decode-phase self-attention, showing that CPUs can sustain reasonable performance on memory-bandwidth-bound attention operations. However, its evaluation assumed substantial CPU resources (e.g., eight 32-core AMD Epyc CPUs for a single A10 GPU) and targeted fixed input/output lengths, limiting its relevance to cost-effective deployments and dynamic online workloads.

MOE-Lightning~\cite{moe-lightning} likewise explored overlapping CPU attention with GPU computation. Its primary focus, however, was offline inference for Mixture-of-Experts (MoE) models, where input and output lengths are typically predetermined and scheduling challenges differ from those in dynamic online serving. As a result, its techniques are less directly applicable to the latency-sensitive, interactive workloads of general-purpose LLM inference.

NEO~\cite{jiang2024neosavinggpumemory} advanced hybrid scheduling by introducing Asymmetric Pipelining and extending consideration to both prefill and decode stages. While NEO demonstrates gains under extreme GPU memory constraints, its scheduler depends on heuristics that can yield suboptimal decisions. More critically, for decode-intensive workloads typical of chat and reasoning applications, NEO often defaults to GPU-only execution or resorts to batch splitting, which introduces inefficiencies and leaves CPU resources underutilized. This creates an open problem in precisely the scenarios where decode phases dominate runtime and sustained performance is most critical.

These limitations motivate the need for a profiling-informed scheduling strategy that dynamically overlaps CPU-offloaded attention with GPU computation in decode-intensive workloads, maximizing parallelism while avoiding performance degradation.

\subsection{Asymmetric Pipelining Challenges in Decode-Intensive Workloads}
\label{sec:motiv}

Among existing hybrid approaches, NEO’s Asymmetric Pipelining is the most advanced attempt at CPU--GPU coordination, yet it exposes key limitations when applied to decode-intensive workloads. To clarify why current solutions fall short and to motivate our design, we analyze these challenges in detail.

The Asymmetric Pipelining technique addresses GPU memory limits by offloading decode-phase attention and KV cache storage to the CPU. As illustrated in Figure~\ref{fig:asymmetric_pipeline}, NEO forms two concurrent sub-batches: one that runs both GPU and CPU requests (prefill, GPU decoding, and CPU decoding) and another that processes only CPU decoding. This strategy seeks to expand effective GPU batch sizes by leveraging abundant CPU memory while retaining compute-intensive operations on the GPU.

\begin{figure}[t]
\centering
        \includegraphics[width=.8\linewidth]{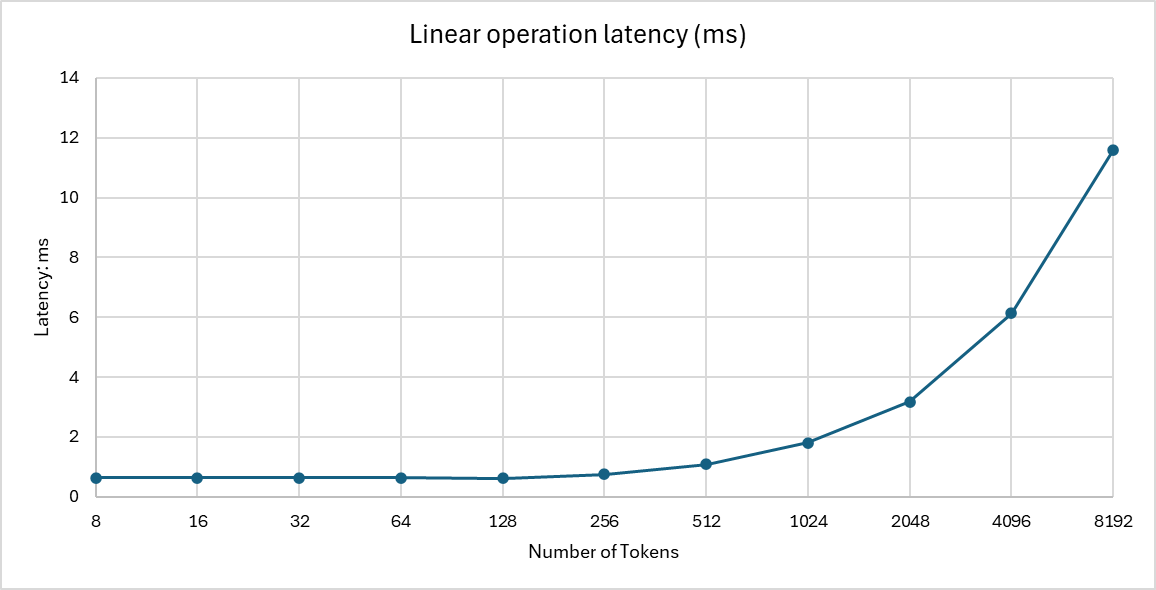}
        \caption{Latency of linear operations (Q, K, V, O projections, FFNs) for one layer of LLaMa-3.1-8B on an A10 GPU, varying with the number of tokens processed.}
        \label{fig:profiling_results}
\end{figure}

NEO's scheduler attempts to balance workloads by enforcing that GPU computation time exceed CPU attention time, ensuring the GPU remains the primary bottleneck rather than the CPU. However, this balancing constraint becomes difficult to satisfy in decode-intensive workloads due to three key factors:




\mytitle{Impact of Batch Splitting on Linear Operations} Our profiling (Figure~\ref{fig:profiling_results}) shows GPU linear operation time remains stable for token counts under 256. In decode phases, where each request processes one token per iteration, typical batch sizes fall within this range. GPUs achieve best utilization when processing decode requests in a single batch. However, Asymmetric Pipelining splits incoming batches into two sub-batches, forcing the GPU to execute linear operations twice per cycle. This duplication often doubles the time spent on linear operations compared to monolithic processing, negating potential parallelism benefits.

\begin{figure}[t]
\centering
        \includegraphics[width=.8\linewidth]{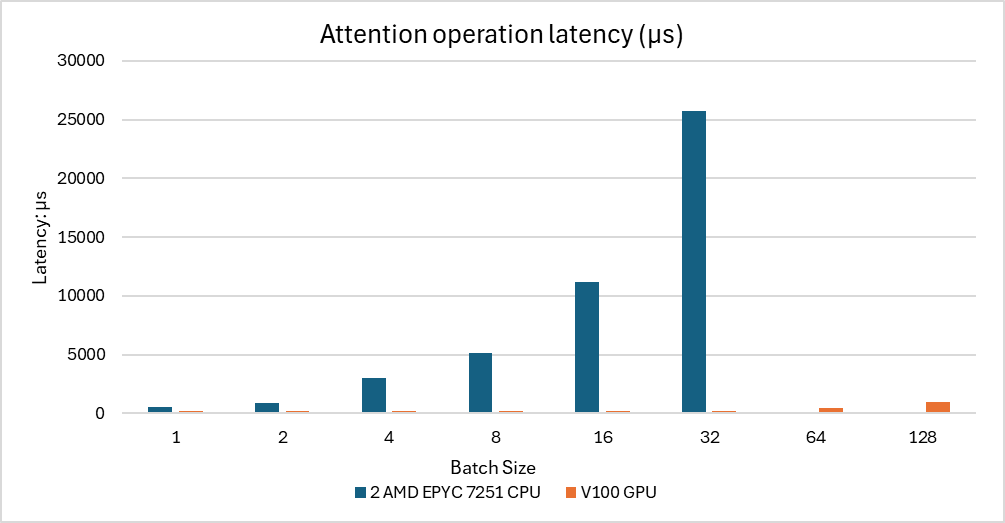}
        \caption{Self-Attention computation latency (hidden size 2048 and sequence length 1024)  on a V100 GPU and two AMD EPYC 7251 CPUs, by batch size. CPU computation time at batch sizes 64 and 128 are not shown due to being substantially higher than GPU latencies.}
        \label{fig:cpu_gpu_attn_comparison}
\end{figure}

\mytitle{CPU-GPU Performance Disparity} CPUs are significantly slower at attention operations than GPUs. Our measurements (Figure~\ref{fig:cpu_gpu_attn_comparison}) show that at batch size 4, CPU attention latency is $3031\mu s$ versus GPU’s $170\mu s$. The gap widens with larger batch sizes, with CPU performance typically under 10\% of GPU throughput. This disparity makes achieving net speedup through Asymmetric Pipelining inherently difficult in decode-only settings, especially when accounting for batch-splitting overheads.

\mytitle{Pipeline Underutilization} NEO’s balancing constraint which requires GPU time to exceed CPU time—severely restricts how many CPU requests can be included in the first sub-batch. As a result, the sub-batch often contains few or no CPU decoding requests, creating pipeline ``bubbles'' where CPU resources remain idle to avoid becoming the system bottleneck, diminishing offloading gains.

These combined effects—doubled linear operation time, pipeline underutilization, and substantial CPU–GPU performance gaps—pose major challenges for Asymmetric Pipelining in decode-intensive workloads. These observations motivate our \emph{Asynchronous Overlap} method (Section~\ref{strategy}), which directly addresses these issues to enable effective CPU–GPU collaboration without the associated penalties.

\section{APEX System Design}
\label{sec:apex_design}

Building on the limitations of existing approaches in decode-intensive scenarios, \textbf{APEX} (Asynchronous Parallel CPU--GPU Execution) introduces a scheduling strategy centered on an \emph{Asynchronous Overlap} mechanism. This section outlines the overall system architecture and the key principles behind this design.

\subsection{System Overview}
\label{sec:system_overview}

\begin{figure}[t]
\centering
    \includegraphics[width=\linewidth]
    {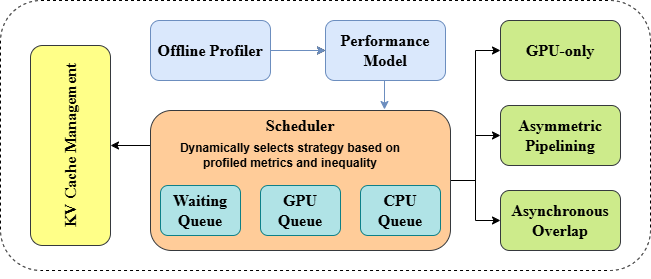}
    \caption{APEX System Architecture}
    \label{fig:system_overview}
\end{figure}

The APEX system architecture, illustrated in Figure~\ref{fig:system_overview}, 
is designed to dynamically manage LLM inference requests by leveraging both GPU and CPU resources. It consists of several key components:

\begin{itemize}
    \item \textbf{Offline Profiler and Performance Model:} Similar to NEO, APEX uses an offline profiler to measure execution times for different model components (e.g., linear layers, GPU/CPU attention) across batch sizes and sequence lengths. Unlike NEO, however, APEX combines these measurements with a runtime-aware performance model that guides fine-grained scheduling decisions without relying solely on heuristics.
    \item \textbf{Request Queues:} Incoming prefill, GPU decode, and CPU decode requests are organized into separate queues.
    \item \textbf{Dynamic Scheduler:} At each iteration, the scheduler selects requests from the queues and, using the performance model together with system state (e.g., GPU memory availability) and request type (prefill vs. decode), determines the optimal execution strategy. Options include GPU-only execution, Asymmetric Pipelining, and our \emph{Asynchronous Overlap} mechanism, which avoids batch splitting.
    \item \textbf{KV Cache Management:} KV cache placement is managed dynamically according to the chosen strategy. Unlike prior systems, APEX reserves GPU memory only for KV pairs of requests actively processed on the GPU, maximizing batching capability without excessive transfers.
\end{itemize}

The core innovation of \textbf{APEX} is its \emph{Asynchronous Overlap} mechanism, designed to improve performance specifically in decode-intensive phases. Importantly, APEX preserves full model accuracy by changing only the execution location of computations, while keeping the mathematical operations and numerical precision identical.

\begin{figure*}[h]
\centering
    \includegraphics[width=\linewidth]
    {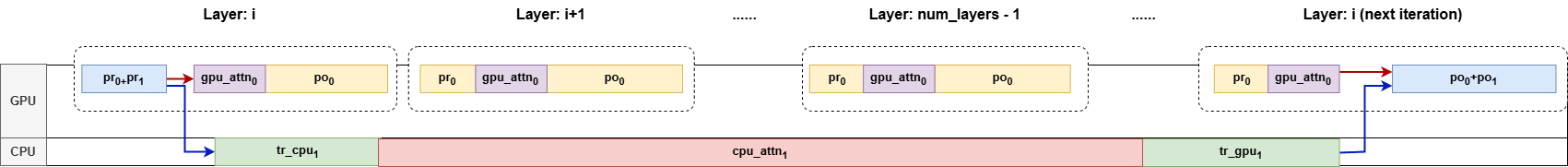}
    \caption{Asynchronous Overlap splits the requests into two sub-batches. The first sub-batch (red arrows) contains prefilling and GPU decoding requests. The second sub-batch (blue arrows) contains only CPU decoding requests. The blue block indicates the synchronization point of GPU/CPU requests. ``pr'' means pre-projection, while “po” means post-projection + FFN operations; “attn” means
attention operations; "tr" means transfer of intermediate value to GPU/CPU.}
    \label{fig:async_overlap}
\end{figure*}

\subsection{Analytical Model for Scheduling Decisions}
\label{sec:analytical_model_for_scheduler}

To determine when offloading decode requests to the CPU via pipelining is beneficial—particularly in deciding whether to apply Asymmetric Pipelining in decode-only scenarios—\textbf{APEX} uses an analytical model. This model guides APEX’s scheduler (Algorithm~\ref{alg:scheduling}) 
in choosing whether conditions favor Asymmetric Pipelining or whether an alternative, such as our proposed \emph{Asynchronous Overlap} mechanism, is more effective.

\mytitle{Parameter Estimation via Offline Profiling} 
The analytical model relies on several key parameters estimated through offline profiling, as described in Section~\ref{sec:system_overview}. Specifically, $T_{glinear}$ and $T_{gatt}$ (GPU computation times for linear operations and attention) are measured across different sequence lengths during the profiling phase. Likewise, CPU attention computation times are measured across both batch size and sequence length. From these measurements, the model derives the maximum number of tokens that can be processed under a given time constraint.

Let $N_G$ denote the rate (tokens per unit time) at which the GPU processes requests, which depends on the total number of tokens in those requests, and let $N_C$ denote the corresponding CPU rate, which depends on both the total number of tokens and the batch size. $T_{glinear}$ represents the GPU computation time for linear operations within a transformer layer for a given batch size in the decode phase, while $T_{gatt}$ denotes the GPU computation time for the self-attention operation for that same batch.

For a GPU-only execution handling a batch of decode requests, the time per iteration ($T_{gpuonly}$) is:
\begin{equation}
    T_{gpuonly} = T_{glinear} + T_{gatt}
\end{equation}

In an Asymmetric Pipelining scenario tailored for decode-only requests, the pipeline structure involves two sub-batches. Due to batch splitting, the linear operations are effectively performed twice in terms of impact on the cycle time. Thus, the effective cycle time ($T_{overlap}$, the total running time of two sub-batches in one transformer layer) is estimated as:
\begin{equation}
    T_{overlap} \approx 2T_{glinear} + T_{gatt}
\end{equation}

Within this pipelined execution, the number of tokens processed by the GPU's attention mechanism in its segment of the pipeline is:
\begin{equation}
    N_{Gtotal} = N_G \times T_{gatt}
\label{eq:gtotal_ap_model}
\end{equation}
The number of tokens processed by the CPU's attention mechanism, operating for the duration $T_{overlap}$ in its segment of the pipeline, is:
\begin{equation}
    N_{Ctotal} = N_C \times T_{overlap} = N_C (2T_{glinear}+T_{gatt})
\label{eq:ctotal_ap_model}
\end{equation}

The throughput for the Asymmetric Pipelining system in this decode-only context is $\frac{N_{Gtotal} + N_{Ctotal}}{T_{overlap}}$. For the GPU-only case, to process a comparable unit of $N_{Gtotal}$ tokens (representing the GPU's attention work), the throughput is $\frac{N_{Gtotal}}{T_{gpuonly}}$. For Asymmetric Pipelining to provide a speedup over a GPU-only approach for this decode workload, its throughput must be higher:
\begin{equation}
    \frac{N_G T_{gatt} + N_C (2T_{glinear} + T_{gatt})}{2T_{glinear} + T_{gatt}} > \frac{N_G T_{gatt}}{T_{glinear} + T_{gatt}}
\label{eq:inequal_throughput_ap_model}
\end{equation}
Performing algebraic transformations on Inequality~\eqref{eq:inequal_throughput_ap_model}, we arrive at the condition required for Asymmetric Pipelining to be beneficial:
\begin{equation}
    \frac{N_G}{N_C} < 2\frac{T_{glinear}}{T_{gatt}} + 3 + \frac{T_{gatt}}{T_{glinear}}
\label{eq:final_condition_ap_model}
\end{equation}

This Inequality~\eqref{eq:final_condition_ap_model} is crucial for APEX's scheduler. It indicates that the effectiveness of Asymmetric Pipelining for decode-only batches depends on the GPU-to-CPU attention speed ratio ($N_G/N_C$) and the system's characteristic ratio of linear operation time to attention time ($T_{glinear}/T_{gatt}$). For typical $T_{gatt}/T_{glinear}$ ratios (between 0.5 and 1.5), $N_G/N_C$ must generally be less than 7.5 to gain the speedup. This implies CPU attention speed ($N_C$) must be at least 13\% of GPU attention speed ($N_G$) for this pipelining strategy to compensate for its overheads.
Given that empirical measurements show $N_C$ is often less than 10\% of $N_G$ (as noted in Section~\ref{sec:motiv}), Inequality~\eqref{eq:final_condition_ap_model} is rarely satisfied in decode-only scenarios. Thus, this model guides the APEX scheduler to employ Asymmetric Pipelining for decode-only batches sparingly, only when the condition holds, and to otherwise choose Asynchronous Overlap.

For mixed workloads involving both prefill and decode requests, APEX extends this analytical framework by accounting for the different computational characteristics of prefill operations. In mixed workload scenarios, the pipeline structure accommodates both prefill and decode phases, leading to a modified cycle time. The effective cycle time becomes:

\begin{equation}
\label{eq:overlap_time}
T_{overlap} = T_{glinear\_pref} + T_{gatt\_pref} + T_{glinear} + T_{gatt}  
\end{equation}

where $T_{glinear\_pref}$ and $T_{gatt\_pref}$ represent the GPU computation times for linear operations and self-attention during the prefill phase, respectively.

Substituting this modified $T_{overlap}$ and running time into the throughput inequality framework established in Equation~\eqref{eq:inequal_throughput_ap_model}, the condition for beneficial Asymmetric Pipelining in mixed workloads becomes:
\begin{equation}
\label{eq:inequal_tp_mixed}
    \frac{N_G T_{gatt} + N_C T_{overlap}}{T_{glinear} + T_{gatt}} > \frac{N_G T_{gatt}}{ T_{gatt}}
\end{equation}

This extended model enables APEX's scheduler to make informed decisions about when to employ Asymmetric Pipelining versus Asynchronous Overlap for mixed workloads, taking into account the additional computational overhead introduced by prefill operations. Since prefill operations typically have longer execution times than decode operations, the extended cycle time often makes the inequality more favorable for Asymmetric Pipelining, as the CPU has more time to perform useful work within each pipeline iteration.

\subsection{Asynchronous Overlap Mechanism}
\label{strategy}

We propose the \emph{Asynchronous Overlap} mechanism to eliminate the need for a separate CPU-only sub-batch in decode-centric workloads and to improve CPU--GPU parallelism. Our approach integrates CPU and GPU decode requests by processing them together in a unified batch during the pre-attention operations (linear layers) of each transformer layer $i$ ($i \in [1,\allowbreak num\_layers]$). This shared GPU execution of linear operations avoids the doubling of $T_{glinear}$ noted in Section~\ref{sec:motiv} 
and leverages the GPU’s efficiency on large, compute-bound batches. After these common linear computations, the execution paths diverge: the query (Q), key (K), and value (V) tensors for requests designated for CPU attention are transferred to CPU memory. The GPU then performs self-attention for its assigned requests while the CPU concurrently executes self-attention for the offloaded ones.

The key innovation in \emph{Asynchronous Overlap} is the strategic scheduling of result synchronization. Instead of immediately transferring CPU-computed self-attention results back to the GPU upon completion (e.g., when the GPU becomes ready for post-attention operations in layer $i$), we deliberately defer this transfer. The results from CPU attention in layer $i$ are only sent back and synchronized just before the GPU requires them to begin the post-attention linear operations of that same layer $i$ in the \textit{subsequent} batch cycle, as illustrated conceptually in Figure~\ref{fig:async_overlap}. 

This design yields two key advantages. First, it maximizes the CPU compute window: delayed synchronization grants the CPU a significantly longer interval to perform attention computations. Unlike Asymmetric Pipelining, where CPU overlap is limited to the GPU’s attention phase, \emph{Asynchronous Overlap} allows the CPU’s compute window for a given layer’s attention to span the GPU’s execution of its pre-attention linear layers, its self-attention across all layers, and its post-attention linear layers within the current iteration. As a result, the CPU effectively has the entire duration of the GPU’s iteration cycle to finish its offloaded attention tasks for a specific layer before synchronization is required in the next cycle.

Second, this approach reduces pipeline bubbles and thereby improves GPU utilization. With a single, strategically deferred synchronization per layer per iteration, the CPU gains a longer and more flexible window to complete its tasks. This design lowers the chance of the CPU becoming a critical-path bottleneck or the GPU stalling while waiting for CPU results. By contrast, experiments with Asymmetric Pipelining show that its two-sub-batch structure can create inefficiencies: the first sub-batch (which may include prefill requests) often contains no CPU-offloaded decode work, producing a “bubble” where CPU resources in that stage remain underutilized. \emph{Asynchronous Overlap} avoids such bubbles by ensuring the CPU always has a continuous and substantial workload derived from the unified batch.

By synchronizing only once per iteration at a non-critical point in the GPU timeline and by processing linear operations on a unified batch, \emph{Asynchronous Overlap} achieves more sustained CPU--GPU parallelism. This approach is especially effective in decode-heavy workloads: even though CPU attention is slower per token than GPU attention, it contributes meaningfully to throughput when given a long, uninterrupted execution window, as this mechanism provides. The scheduler (Section~\ref{algorithm}) 
selects Asynchronous Overlap when Inequality~\eqref{eq:inequal_throughput_ap_model} 
indicates that Asymmetric Pipelining would not be beneficial for the current batch of decode-only requests.

\subsection{Scheduling Algorithm}
\label{algorithm}

The scheduler determines which requests to execute in each iteration and selects the appropriate execution strategy. Our scheduling algorithm is guided by several key principles aimed at optimizing the balance between GPU and CPU processing:

The scheduler follows four primary rules:

\begin{itemize}
    \item \textbf{GPU-first approach:} Consider CPU involvement only if GPU memory cannot hold KV caches for all new requests. When the GPU can handle all requests simultaneously, there is no benefit in involving the CPU due to the GPU's substantially higher computing power.
    
    \item \textbf{Decode-only optimization:} With only decoding requests present, the scheduler evaluates inequality~\eqref{eq:final_condition_ap_model} to determine whether asymmetric pipelining would provide speedup compared to a GPU-only strategy. At this point, GPU memory is typically fully utilized, so $T_{glinear}$ and $T_{gatt}$ remain stable. The scheduler need only calculate how many tokens the CPU can process within the time window $2T_{glinear}+T_{gatt}$. If the inequality holds, the scheduler selects asymmetric pipelining; otherwise, it switches to asynchronous overlap.
    
    \item \textbf{Mixed workload handling:} With both prefilling and decoding requests present, the scheduler applies a modified version of the inequality. Although prefilling requests exist, the comparison still focuses on CPU decoding versus GPU decoding speeds, as prefilling and decoding have different per-token processing times that cannot be directly compared. The values $N_G$, $T_{gatt}$, and $T_{glinear}$ remain unchanged since GPU decoding speed is fixed. However, the CPU running time increases as shown in equation~\ref{eq:overlap_time}. We use use inequality~\ref{eq:inequal_tp_mixed} to decide which strategies to use. In this scenario, the CPU has more time to process tokens, making speedup of asymmetric pipelining more achievable.

    \item \textbf{Partial progress prioritization:} When using asynchronous overlap, if some CPU requests have already completed the first $i$ layers of decoding when new prefilling requests arrive, and the inequality holds, the scheduler prioritizes these partially processed CPU requests in the decode-only sub-batch if the other sub-batch is full. This optimization is valuable because these requests add only $(num\_layers-i) \cdot T_{glinear}$ extra time instead of $num\_layers \cdot T_{glinear}$ extra time.
\end{itemize}

\begin{algorithm}[h!]
\caption{APEX Main Scheduler}
\label{alg:scheduling}
\begin{algorithmic}
\STATE {\bfseries Input:} Prefill queue $P_{in}$, GPU decode queue $D_{gpu\_in}$, CPU decode queue $D_{cpu\_in}$
\STATE {\bfseries Output:} Execution strategy 

\IF{$D_{cpu\_in}$ is empty}
    \STATE // GPU has available memory
    \STATE {\bfseries return} GPU-only strategy 
\ENDIF

\STATE // \textsc{AsymPipeStrat}: selects candidate sub-batches for Asymmetric Pipelining given current queues/memory
\STATE $(P_{sch}, D_{gpu\_sch}, D_{cpu\_sch}) \leftarrow$ AsymPipeStrat($P_{in}$, $D_{gpu\_in}$, $D_{cpu\_in}$)

\IF {$D_{cpu\_sch}$ is empty}
    \STATE //No CPU requests can be effectively pipelined
    \STATE {\bfseries return} Asynchronous Overlap strategy
\ENDIF

 \STATE {\bfseries return} StrategySelection($P_{sch}$, $D_{gpu\_sch}$, $D_{cpu\_sch}$)

\end{algorithmic}
\end{algorithm}

\begin{algorithm}[h!]
\caption{APEX Strategy Selection}
\label{alg:selection}
\begin{algorithmic}
\STATE {\bfseries Function} StrategySelection($P_{gpu}$, $D_{gpu}$, $D_{cpu}$):

\STATE $T_{glinear}, T_{gatt} \leftarrow$ Get profiled GPU times (decode)

\IF{$P_{gpu}$ is empty}
    \STATE // Decode-only scenario 
    \STATE $T_{overlap}\leftarrow 2T_{glinear} + T_{gatt}$
    \STATE $throughput_{gpu} \leftarrow \frac{N_GT_{gatt}}{T_{glinear} + T_{gatt}}$
    \STATE $throughput_{pipe} \leftarrow \frac{N_GT_{gatt} + N_C \cdot T_{overlap}}{2T_{glinear} + T_{gatt}}$
    
\ELSE
    \STATE // Mixed workload scenario  
    \STATE $T_{glinear\_pref}, T_{gatt\_pref} \leftarrow$ Get profiled GPU times (prefill)
    \STATE $T_{overlap} \leftarrow T_{glinear\_pref} + T_{gatt\_pref} + T_{glinear} + T_{gatt}$

     \STATE $throughput_{gpu} \leftarrow \frac{N_GT_{gatt}}{T_{gatt}}$
    \STATE $throughput_{pipe} \leftarrow \frac{N_GT_{gatt} + N_C \cdot T_{overlap}}{T_{glinear} + T_{gatt}}$

\ENDIF

\IF {$throughput_{pipe} > throughput_{gpu}$}
    \STATE {\bfseries return} Asymmetric Pipelining strategy
\ELSE
    \STATE {\bfseries return} Asynchronous Overlap strategy
\ENDIF

\end{algorithmic}
\end{algorithm}

The APEX scheduling procedure, shown in Algorithms~\ref{alg:scheduling} and \ref{alg:selection}, adapts dynamically to workload characteristics by first attempting Asymmetric Pipelining partitioning and then evaluating its potential speedup using the analytical model described in Section~\ref{sec:analytical_model_for_scheduler}. Algorithm~\ref{alg:scheduling} captures the main scheduling logic for request queue management and for deciding when CPU involvement is beneficial, while Algorithm~\ref{alg:selection} applies the throughput-based model to compare Asymmetric Pipelining against GPU-only execution. This two-stage design optimizes resource allocation across GPU and CPU to maximize throughput while preserving latency, ensuring hybrid execution is invoked only when it provides measurable gains over pure GPU processing.

A key element of robustness—particularly under the \emph{Asynchronous Overlap} strategy—is CPU--GPU synchronization. Before the designated synchronization point (i.e., just prior to the GPU initiating post-attention operations for a given layer), the GPU checks whether the CPU’s offloaded computation for that layer has completed. If the results are not yet available, the GPU does not stall; it continues executing its current tasks for the iteration. The readiness check is repeated in the next iteration, and synchronization proceeds only once the CPU results are available.

\section{Implementation}
We implement APEX by extending the open-source NEO system.
Our implementation introduces two key contributions: (1) a new CPU paged-attention backend built on Llamafile, and (2) a multi-threaded runtime using Pybind11 to enable asynchronous GPU--CPU overlap.

\subsection{Integration of Llamafile Kernels}
NEO implements a custom paged-attention kernel for CPU with ISPC, a language for writing SPMD programs on CPUs. In contrast, we implement the CPU paged-attention kernel using the Llamafile matrix-multiplication kernels. Llamafile is a Mozilla open-source project that combines \texttt{llama.cpp} with Cosmopolitan Libc into a single framework, producing a single-file executable (a ``llamafile'') that runs locally on most computers without installation. It provides efficient CPU matrix-multiplication kernels adopted by \texttt{llama.cpp} and \texttt{ktransformers}. We implement our CPU paged-attention using the Llamafile kernels. In this setting, CPU paged-attention is slightly slower at small batch sizes than NEO's custom kernel; however, as batch size increases, our implementation achieves up to a $2\times$ speedup over NEO's custom paged attention. This improved CPU performance substantially increases opportunities to overlap CPU paged-attention with concurrent GPU operations. In addition, we pin CPU memory for the KV cache to accelerate transfers between GPU and CPU. 

Unlike NEO’s ISPC-based CPU attention, our Llamafile backend couples high-throughput MM kernels with pinned host KV buffers, reducing host–device copy latency. This backend is optimized for the unified-batch execution path (Section~\ref{strategy}), which NEO does not implement due to its sub-batch design.

\subsection{Asynchronous GPU--CPU Overlap Runtime}
\label{sec:overhead}
To realize asynchronous overlap between CPU and GPU operations, our system employs a two-thread Python architecture. The main thread manages core system logic and launches CUDA kernels for GPU tasks, while a dedicated compute thread executes CPU-bound attention. To enable true parallelism in the presence of Python's Global Interpreter Lock (GIL), we integrate C++ routines for attention via Pybind11 and use \texttt{gil\_scoped\_release} so the GIL is released during C++ execution, allowing the main thread to dispatch GPU work concurrently. While this approach enables the desired overlap, Python thread contention and scheduling overheads are inherent. Consequently, the CPU compute thread must process a sufficient number of requests per iteration to amortize these overheads. We therefore enforce a minimum threshold for the number of CPU requests required to initiate CPU-based attention. Our empirical evaluation indicates these overheads are adequately offset when the number of CPU requests is at least eight times the number of GPU requests.

Unlike NEO’s two-sub-batch pipeline, APEX executes linear layers once per iteration on a single unified GPU batch, then branches attention: GPU-attended requests remain on device, while offloaded requests transfer Q/K/V to host.

\section{Evaluation}
\subsection{Experiment Setup}

\mytitle{Testbed} We run experiments on two systems: (1) dual Intel Xeon Gold 6342 processors (2.80~GHz, 24 cores per processor) with an NVIDIA A10 GPU, and (2) dual Intel Xeon Gold 6130 processors (2.10~GHz, 16 cores per processor) with an NVIDIA T4 GPU. We set $max\_num\_batched\_tokens$ to 800 and 20{,}000 on T4 and A10, respectively. The hardware characteristics of our experimental settings are shown in Table~\ref{tab:hw}.

\begin{table}[htbp]
\caption{Hardware settings of our experiments}
\begin{center}
\begin{tabular}{|c|c|c|}
\hline
\textbf{CPU (cores)} & \textbf{RAM} & \textbf{GPU} \\
\hline
2$\times$ Xeon Gold 6342 (2$\times$24 cores) & 250~GiB & NVIDIA A10 \\ 
\hline
2$\times$ Xeon Gold 6130 (2$\times$16 cores) & 180~GiB & NVIDIA T4 \\
\hline
\end{tabular}
\label{tab:hw}
\end{center}
\end{table}

\mytitle{Models} We evaluate LLaMa-2-7B on the T4 platform and LLaMa-3.1-8B on the A10 platform. These pairings ensure compatibility between model size and GPU memory (LLaMa-2-7B with the T4’s 16~GB memory; LLaMa-3.1-8B with the A10’s 24~GB memory) and reflect realistic deployment scenarios shaped by resource constraints and performance requirements.

\mytitle{Baseline}
We compare APEX against two baselines:
\begin{itemize}
    \item \textbf{vLLM}~\cite{kwon2023efficient}, a state-of-the-art high-throughput and memory-efficient GPU-only inference and serving engine for LLMs, designed primarily for high-end GPUs. 
    \item \textbf{NEO}~\cite{jiang2024neosavinggpumemory}, a GPU-CPU hybrid inference engine that mitigates GPU memory pressure via CPU offloading. APEX extends NEO’s framework but introduces unified-batch execution, delayed synchronization, and model-driven scheduling for improved performance on decode-heavy workloads.
\end{itemize}

These baselines capture both ends of the design space: vLLM represents the state-of-the-art GPU-only inference engine, while NEO represents the leading hybrid CPU--GPU approach most similar to our setting. Other systems (e.g., FlexGen~\cite{sheng2023flexgenhighthroughputgenerativeinference}, FastDecode~\cite{he2024fastdecodehighthroughputgpuefficientllm}, PowerInfer~\cite{song2023powerinfer}) target different usage scenarios (offline batch inference, A100-class GPUs, or CPU-only execution) and are therefore less directly comparable to our online, memory-constrained setting.

\mytitle{Workloads} We evaluate APEX using several real-world workloads covering different task types:
\begin{itemize}
    \item \textbf{Azure LLM inference trace (AZ)}~\cite{stojkovic2024dynamollmdesigningllminference} is a sample from multiple LLM inference services in Azure, collected between May 10--19, 2024, and reflects realistic multi-tenant conversational workloads. 
    \item \textbf{LiveBench (LB)}~\cite{white2025livebenchchallengingcontaminationlimitedllm} is a benchmark designed to mitigate test-set contamination and provide objective evaluation across diverse LLM tasks. 
    \item \textbf{Dolphin-r1 (Dolphin)}~\cite{quixiai2024dolphinr1} is an 800k-sample dataset with composition similar to the data used for training DeepSeek-R1 Distill models, representing long-form reasoning and instruction-following. 
    \item \textbf{OpenAI Summarization Comparison (OSC)}~\cite{carperai2024openaisummarizecomparisons} consists of input texts paired with human-selected (``chosen'') and rejected summaries, drawn from real-world human--chatbot interactions, making it representative of practical summarization workloads.
\end{itemize}

\subsection{Throughput}

\begin{figure*}[h!]
    \centering
    \begin{subfigure}[b]{0.45\textwidth}
        \centering
        \includegraphics[width=\textwidth]{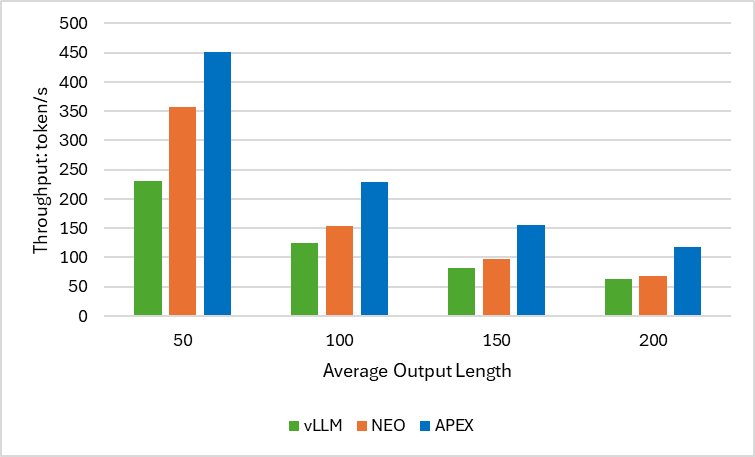}
        \caption{T4 + LLaMa-2-7B + OSC with varied average output length}
        \label{fig:fig3}
    \end{subfigure}
    \hfill
    \begin{subfigure}[b]{0.45\textwidth}
        \centering
        \includegraphics[width=\textwidth]{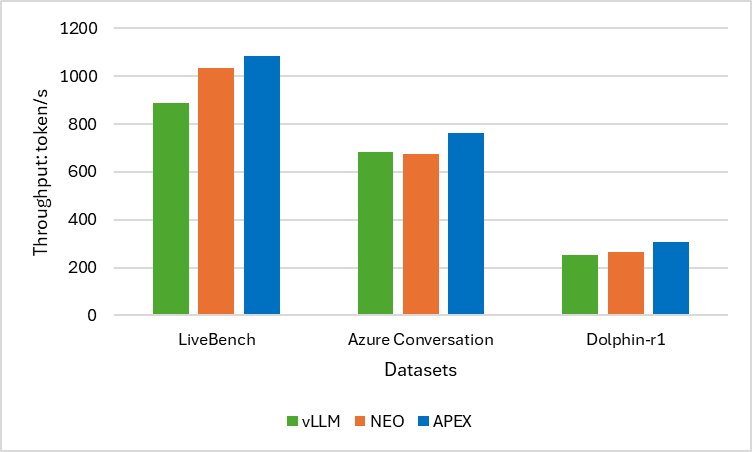}
        \caption{A10 + LLaMa-3.1-8B + 3 workloads (AZ, LB, Dolphin)}
        \label{fig:fig4}
    \end{subfigure}
    \caption{Throughput in different GPUs and different synthetic workloads }
    \label{fig:tp}
\end{figure*}

\begin{figure*}[t]
    \centering
    \begin{subfigure}[b]{0.32\textwidth}
         \centering
        \includegraphics[width=\textwidth]{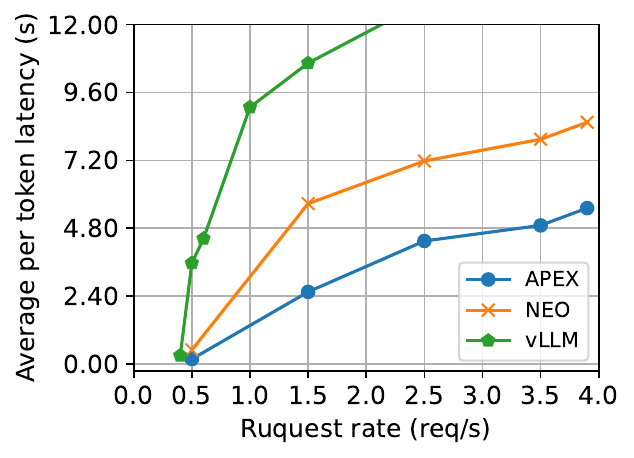}
        \caption{T4 + LLaMa-2-7B + OSC}
        \label{fig:fig1}
    \end{subfigure}
    \hfill
    \begin{subfigure}[b]{0.32\textwidth}
        \centering
        \includegraphics[width=\textwidth]{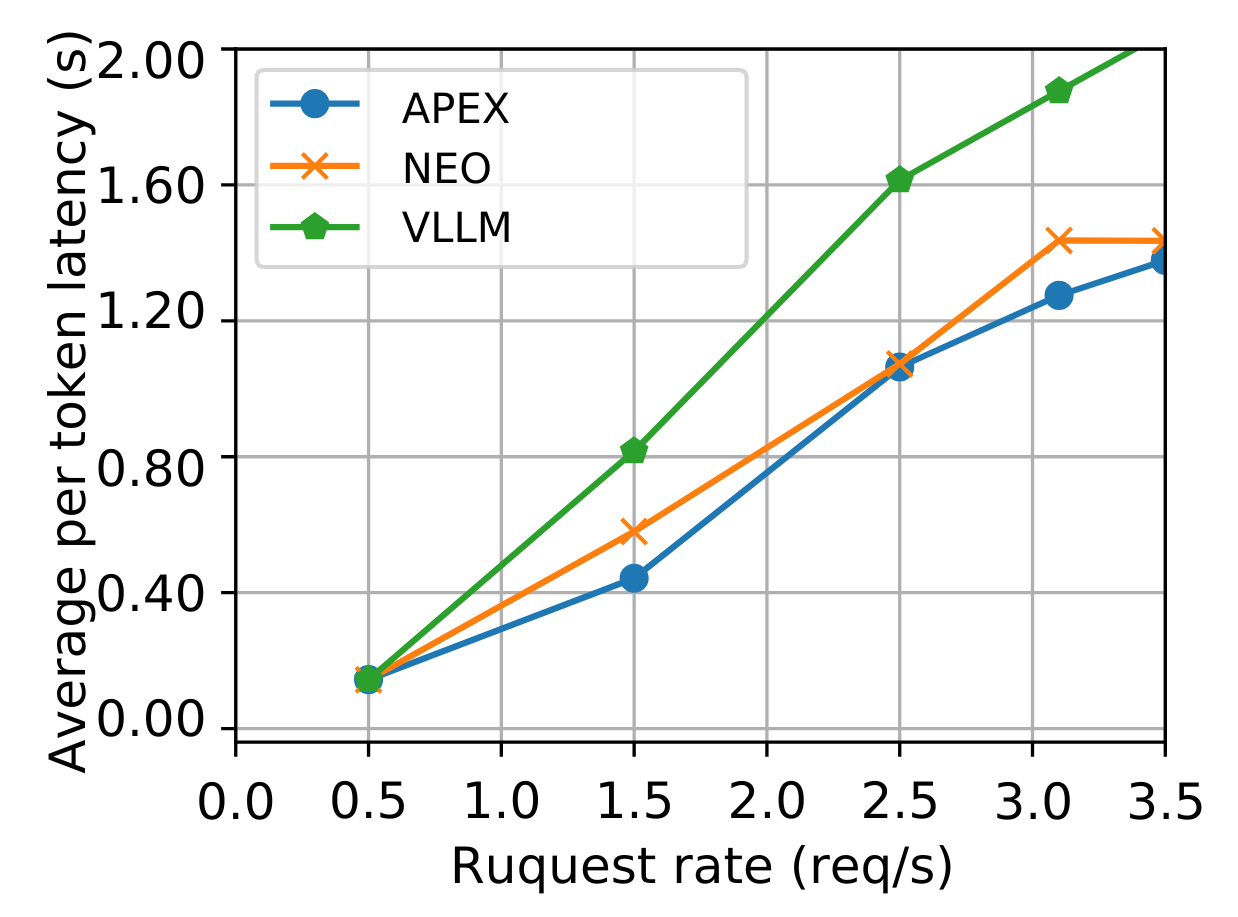}
        \caption{A10 + LLaMa-3.1-8B + AZ}
        \label{fig:fig2}
    \end{subfigure}
    \hfill
    \begin{minipage}[b]{0.32\textwidth}
        \centering
        \includegraphics[width=\textwidth]{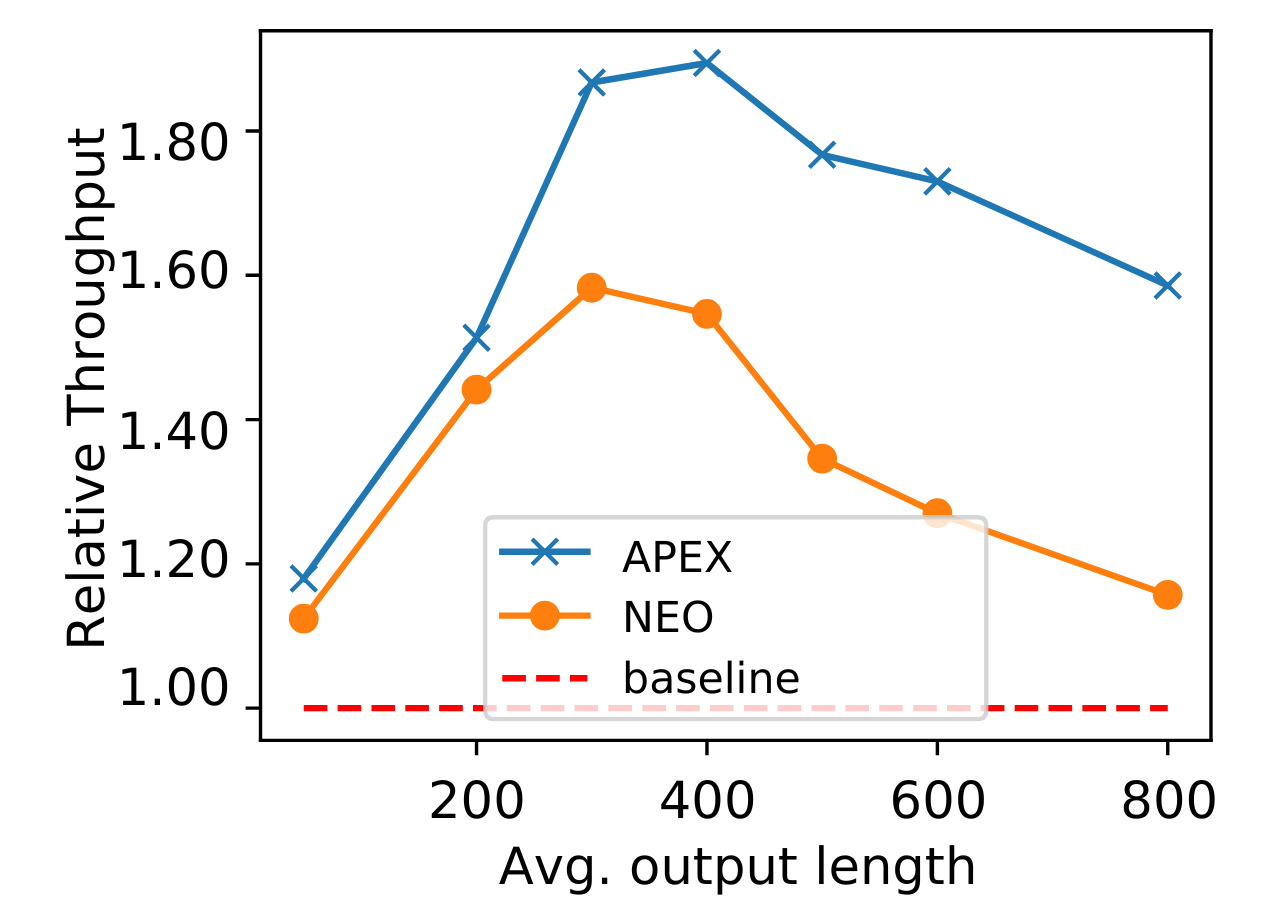}
    \end{minipage}
    
    \vspace{0.3cm}
    
    \begin{minipage}{0.65\textwidth}
        \caption{Average per token latency curve comparison between NEO, vLLM and APEX on T4 and A10 GPUs. For each request, we compute its per-token latency by dividing its full latency by its output token number, and then we take the average among all requests.}
        \label{fig:lat}
    \end{minipage}
    \hfill
    \begin{minipage}{0.32\textwidth}
        \caption{Relative throughput in varied average output length compared with the GPU-only baseline. The average input length is 1000.}
        \label{fig:length}
    \end{minipage}
\end{figure*}
We evaluated APEX throughput against baseline systems on both A10 and T4 GPU architectures, with results shown in Figure~\ref{fig:tp}. On the A10 GPU, APEX achieved consistent improvements across three benchmark workloads: LiveBench (6\% over NEO, 22\% over vLLM), Azure Conversation (13\% over NEO, 11\% over vLLM), and Dolphin-r1 (16\% over NEO, 20\% over vLLM). These results demonstrate that APEX delivers robust throughput gains across workloads with diverse characteristics.

For the T4 GPU evaluation, we employed the OSC workload and varied the average output length. Under this configuration, APEX achieved significantly higher throughput, outperforming NEO by up to 72\% and vLLM by up to 96\%.
The observed speedup primarily stems from APEX’s ability to overlap CPU and GPU computations during decode-intensive phases of request processing. The magnitude of this gain is influenced by two principal factors: (i) the computational power disparity between GPU and CPU, and (ii) the fraction of total execution time dominated by decode-intensive phases. Formally, if we denote the ratio of CPU to GPU computational power as $\rho_c$ and the fraction of decode-intensive time to total execution time as $\rho_t$, the achievable speedup $S$ can be approximated as:
$S \approx \rho_c \cdot \rho_t$
\footnote{This relation follows from hybrid execution where decode operations (occupying $\rho_t$ of the workload) are parallelized across GPU and CPU, with the CPU operating at $\rho_c$ times the GPU’s speed. Under ideal load balancing, the net speedup is bounded by the effective utilization of both processors, yielding $S \approx \rho_c \rho_t$.}.

The larger throughput gains observed on the T4 platform, compared to the A10, can be explained by these factors. The T4 GPU provides considerably lower computational power than the A10, while CPU processing capabilities are comparable across both setups, yielding a larger $\rho_c$ in the T4 experiments. In addition, the OSC workload used on the T4 features a greater fraction of decode-intensive time (larger $\rho_t$) than the workloads evaluated on the A10. Consequently, the combination of a less powerful GPU (higher $\rho_c$) and a workload with stronger decode dominance (higher $\rho_t$) together accounts for the more pronounced speedup achieved by APEX on the T4 GPU.

\subsection{Average Per-Token Latency}
In addition to throughput, we evaluated the average per-token latency (including queuing time) for APEX compared to vLLM and NEO on both T4 and A10 GPUs. The results, shown in Figure~\ref{fig:lat}, highlight clear latency advantages for APEX. On the T4 GPU, APEX reduced average per-token latency to nearly half that of NEO. On the A10 GPU, APEX also achieved slightly lower latency than NEO.

Taken together with the throughput results, these findings demonstrate that APEX not only delivers higher throughput than NEO but also sustains comparable or better latency performance.


\subsection{Varying Output Lengths}
To further examine the performance characteristics of APEX in relation to sequence length, we conducted an evaluation on the A10 GPU. Throughput values of NEO and APEX were normalized to a consistent GPU-only baseline. The average output length was systematically varied, and the results, shown in Figure~\ref{fig:length}, align closely with our theoretical framework of CPU-GPU computation overlap.

The data reveals distinct performance regimes. For shorter outputs (50--200 tokens), APEX achieves a modest 5\% throughput improvement over NEO. In this range, execution time is strongly influenced by the prefill stage and a mixture of prefill and decode operations, leaving fewer opportunities for APEX to exploit CPU-GPU parallelism. As average output length increases (200--500 tokens), the performance gap widens (17\%, 22\%, 31\%), reflecting the growing dominance of the decode phase, where APEX excels. Beyond 500 tokens, decode operations constitute the majority of processing time, and APEX reaches its maximum observed benefit, achieving up to 37\% higher throughput than NEO.

For very long outputs (above 600 tokens), the throughput advantage of APEX over NEO stabilizes. This plateau effect is consistent with our model: once the proportion of decode-intensive time ($\rho_t$) saturates, the achievable speedup is bounded by the CPU-to-GPU compute power ratio ($\rho_c$). Thus, while APEX substantially accelerates decode-heavy workloads, its ultimate performance ceiling is dictated by fundamental hardware balance.

    
    


\subsection{Ablation Studies}

We evaluate the individual contributions of APEX’s three key innovations: the Asynchronous Overlap mechanism (AO), the custom CPU Paged Attention kernel (AK), and analytical modeling (AM). Experiments were conducted with three configurations—AO, AO+AK, and AO+AK+AM—each compared against NEO to isolate the speedup from each component.

\mytitle{Experimental Setup} We evaluate these configurations on a T4 GPU using the OSC dataset with varying average output lengths, measuring incremental performance gains from each innovation.

\begin{figure}
\centering
    \includegraphics[width=\linewidth]
    {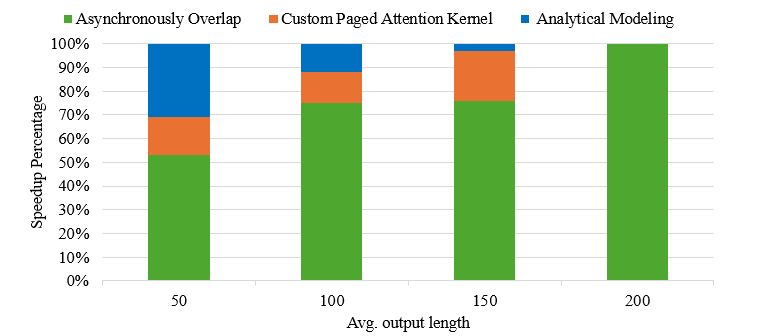}
    \caption{Speedup Contribution of APEX Components by Average Output Length. The average input length is 1000.}
    \label{fig:ablation}
\end{figure}

\mytitle{Results} Figure~\ref{fig:ablation} reports the speedup achieved by each configuration. The Asynchronous Overlap mechanism consistently provides substantial gains (53--100\%) across all settings, validating our central hypothesis that effective CPU-GPU overlap significantly improves throughput regardless of sequence length.

The custom CPU Paged Attention kernel contributes an additional 13--21\% speedup in most cases, though its benefit diminishes for longer outputs (e.g., negligible at 200 tokens). In these cases, longer sequences allow the CPU to finish requests within a single iteration even using NEO’s kernel, reducing the marginal impact of kernel-level optimizations.

The analytical modeling component yields the highest contribution at shorter lengths (up to 31\% speedup) but declines with longer outputs, becoming negligible by 200 tokens. This occurs because the model compares Asymmetric Pipelining against GPU-only execution. At long outputs, Asymmetric Pipelining often fails to schedule CPU requests at all, bypassing the analytical decision step and defaulting directly to Asynchronous Overlap.

Overall, these results confirm that while each innovation contributes meaningfully, the Asynchronous Overlap mechanism is the dominant source of APEX’s performance improvements, delivering consistent and substantial gains across diverse workload characteristics.

\section{Discussion}
\mytitle{Comparison with NEO} 
APEX builds upon the Asymmetric Pipelining technique originally proposed by NEO but enhances its effectiveness through a more selective activation strategy. The key distinction is APEX’s use of an analytically derived inequality criterion to determine precisely when Asymmetric Pipelining is beneficial for a given workload. This criterion has proven more accurate at predicting real speedup than approaches based on coarse metrics such as aggregate requests per second. By targeting only favorable conditions, APEX maximizes the benefits of Asymmetric Pipelining while minimizing cases where its use would lead to slowdowns. Moreover, when the criterion indicates limited benefit, APEX seamlessly switches to its alternative optimization, Asynchronous Overlap, ensuring continued performance gains even in scenarios where NEO defaults to less efficient execution.

\mytitle{CPU Attention Task Pool for Enhanced Flexibility} 
Currently, APEX processes CPU attention requests in batches, handling all requests for a single layer within one iteration. Looking forward, we propose a CPU attention computation task pool to allow finer-grained scheduling. As new requests arrive, they could be dynamically added to this pool, with the CPU servicing tasks according to a layer-wise prioritization scheme (e.g., processing from later layers first, or by heuristics based on computational cost or data dependencies). This approach would enable the CPU to process tasks from multiple layers within a single iteration, reducing idle time and improving overall utilization. The trade-off is that such advanced scheduling would require additional memory to store intermediate values (e.g., residuals, outputs) from multiple layers, increasing both engineering complexity and memory footprint.

\section{Related Work}

\mytitle{Evolution of Offloading Approaches} 
Early approaches to offloading for LLM inference primarily sought to overcome GPU memory limits by moving model parameters or entire KV caches to CPU memory or even disk storage. FlexGen~\cite{sheng2023flexgenhighthroughputgenerativeinference} pioneered dynamic offloading across a memory hierarchy, demonstrating feasibility for models exceeding GPU capacity. Similarly, llama.cpp~\cite{githubGitHubGgmlorgllamacpp} enables execution on CPU-centric or SSD-assisted setups. However, when applied to the frequently accessed KV cache during the decode phase, such strategies often incur prohibitive latencies due to repetitive transfers over PCIe, making them unsuitable for low-latency online services~\cite{sheng2023flexgenhighthroughputgenerativeinference}. Efforts to mitigate this through algorithmic I/O reduction, such as sparsity-based selective loading in LLMFlash~\cite{alizadeh2024llmflashefficientlarge} and PowerInfer~\cite{song2023powerinfer, xue2024powerinfer2fastlargelanguage}, provide partial relief but are often model-dependent and do not address the core challenge of actively coordinating computations across heterogeneous processors during the decode phase.

\mytitle{Hybrid Execution Architectures} 
Recognizing these bottlenecks, hybrid execution has emerged as a more targeted strategy for the decode-phase attention mechanism. This approach offloads KV cache storage and attention computation to the CPU while retaining compute-intensive MLP layers on the GPU. Its feasibility has been demonstrated in several systems, as discussed in Section~\ref{subsec:hybrid_inference_background}. Beyond CPU-GPU hybrids, broader exploration of inference optimization on diverse hardware—including accelerators and Processing-in-Memory (PIM) devices~\cite{tpu, pim, ortega2024pimainovelarchitecturehighefficiency, llmpq, patel2024splitwiseefficientgenerativellm}—underscores the need for intelligent workload management in heterogeneous environments.

\section{Conclusion}
We presented APEX, a performance-model-driven scheduling strategy designed to maximize CPU-GPU parallelism for online Large Language Model (LLM) inference on memory-constrained GPUs. APEX predicts CPU and GPU subtask execution times to dynamically dispatch computations and introduces an Asynchronous Overlap mechanism that hides CPU latency without incurring batch-splitting overheads. We evaluated APEX using LLaMa-2-7B and LLaMa-3.1-8B on NVIDIA T4 and A10 GPUs. Results show that APEX improves throughput by 84\%--96\% on T4 and 11\%--89\% on A10 over the GPU-only vLLM, while achieving up to 49\% (T4) and 37\% (A10) higher throughput than the state-of-the-art hybrid scheduler NEO in long-output settings, all without compromising average per-token latency. These results demonstrate that APEX enables efficient real-time LLM inference on edge and mid-range GPUs, providing a blueprint for future heterogeneous scheduling frameworks.

\bibliographystyle{ACM-Reference-Format}
\bibliography{sample-base}

\end{document}